\newif\ifNotes\Notesfalse
\newif\ifAnon\Anonfalse
\newif\ifCamera\Camerafalse

\ifCamera
  \Notesfalse
  \Anonfalse
\fi

\documentclass[conference,final,10pt]{IEEEtran}
\usepackage{cite}
\usepackage{amsmath,amssymb,amsfonts}
\usepackage{algorithm}
\usepackage{algpseudocode} 
\usepackage{graphicx}
\usepackage{textcomp}
\usepackage{xcolor}
\usepackage{fancyhdr}
\usepackage[hyphens]{url}
\usepackage[numbers]{natbib}
\usepackage{hyperref}
\usepackage{tabu}
\usepackage{xspace}
\usepackage{enumitem}
\usepackage[nameinlink,capitalise]{cleveref}
\usepackage{pgfplots}
\usepackage{tikz}
\usepackage{tabularx}
\usepackage{colortbl}
\usepackage{multirow,array}
\usepackage{booktabs}
\usetikzlibrary{arrows}
\usetikzlibrary{shadows}
\usepackage{balance}

\def\algbackskip{\hskip-\ALG@thistlm}

\hypersetup{
        colorlinks,
        linkcolor={red!50!black},
        citecolor={blue!80!black},
	urlcolor={blue!80!black}}

\def\BibTeX{{\rm B\kern-.05em{\sc i\kern-.025em b}\kern-.08em
    T\kern-.1667em\lower.7ex\hbox{E}\kern-.125emX}}

\ifNotes
\newcommand{\colorcomment}[2]{\leavevmode\unskip\space{\color{#1}#2}\xspace}
\else
\newcommand{\colorcomment}[2]{\leavevmode\unskip\relax}
\fi

\definecolor{darkgreen}{rgb}{0,0.65,0}
\newcommand{\taggedcolorcomment}[3]{\colorcomment{#1}{[\textbf{#2}: #3]}}

\newcommand{\thomas}[1]{\taggedcolorcomment{darkgreen}{thomas}{#1}}
\newcommand{\austin}[1]{\taggedcolorcomment{blue}{austin}{#1}}

\crefname{figure}{Figure}{Figures}

\newcommand{\cc}{\textsf{\small{Chameleon Cache}}\xspace}

\newcommand{\rsclp}{\textit{Randomized Skewed Caches}\xspace}   
\newcommand{\rscss}{RSC\xspace}  
\newcommand{\rscsp}{RSCs\xspace} 
\newcommand{\vc}{VC\xspace}

\newcommand{\ppp}{\textsf{\small Prime+\allowbreak Prune+\allowbreak Probe}\xspace}

\newcommand{\soaddress}{second-\allowbreak order address\xspace}
\newcommand{\soaddresses}{second-\allowbreak order addresses\xspace}

\definecolor{grayy}{HTML}{626262}
\newcommand{\RC}[1]{\cellcolor{grayy!#1}}

\title{Chameleon Cache: Approximating Fully Associative Caches with Random Replacement to Prevent Contention-Based Cache Attacks} 
\ifAnon
  \author{}
\else
\author{
  \IEEEauthorblockN{Thomas Unterluggauer\IEEEauthorrefmark{1}, Austin Harris\IEEEauthorrefmark{2}, Scott Constable\IEEEauthorrefmark{1}, Fangfei Liu\IEEEauthorrefmark{1}, Carlos Rozas\IEEEauthorrefmark{1} }
\vspace{0.05in}
\IEEEauthorblockN{\IEEEauthorrefmark{1}Intel Corporation\ \emph{\{thomas.unterluggauer,scott.d.constable,fangfei.liu,carlos.v.rozas\}@intel.com}}
\IEEEauthorblockN{\IEEEauthorrefmark{2}UT Austin\ \emph{austinharris@utexas.edu}}
}
\fi

\begin{document}
\maketitle
\pagestyle{empty}
\renewcommand{\headrulewidth}{0pt}
\thispagestyle{fancy}

\fancyhead{}\fancyfoot{}

\fancyfoot[C]{\small© 2022 IEEE.  Personal use of this material is permitted.  Permission from IEEE must be obtained for all other uses, in any current or future media, including reprinting/republishing this material for advertising or promotional purposes, creating new collective works, for resale or redistribution to servers or lists, or reuse of any copyrighted component of this work in other works. DOI 10.1109/SEED55351.2022.00009}



\begin{abstract} 
Randomized, skewed caches (RSCs) such as CEASER-S have recently received much attention to defend against contention-based cache side channels. By randomizing and regularly changing the mapping(s) of addresses to cache sets, these techniques are designed to obfuscate the leakage of memory access patterns. However, new attack techniques, e.g., \ppp, soon demonstrated the limits of RSCs as they allow attackers to more quickly learn which addresses contend in the cache and use this information to circumvent the randomization. To yet maintain side-channel resilience, RSCs must change the random mapping(s) more frequently with adverse effects on performance and implementation complexity. 

This work aims to make randomization-based approaches more robust to allow for reduced re-keying rates and presents \cc. \cc extends RSCs with a victim cache (VC) to decouple contention in the RSC from evictions observed by the user. The VC allows \cc to make additional use of the multiple mappings RSCs provide to translate addresses to cache set indices: when a cache line is evicted from the RSC to the VC under one of its mappings, the VC automatically reinserts this evicted line back into the RSC by using a different mapping. As a result, the effects of previous RSC set contention are hidden and \cc exhibits side-channel resistance and eviction patterns similar to fully associative caches with random replacement. We show that \cc has performance overheads of $<1\%$ and stress that VCs are more generically helpful to increase side-channel resistance and re-keying intervals of randomized caches.
\end{abstract}

\section{Introduction}

Cache side channels have been intensively studied over the past two decades as these allow to circumvent architectural isolation boundaries and reveal sensitive information being processed by applications running on the same system. Over time, the scope of cache side channels has expanded from cryptographic targets \cite{OsvikST06,GullaschBK11,LiuYGHL15,BruinderinkHLY16} to other domains such as AI \cite{YanFT20} and the more recent transient execution attacks (e.g., Spectre\cite{KocherHFGGHHLM019}, Meltdown \cite{Lipp0G0HFHMKGYH18}) and thus sparked interest in potential mitigations.

Fundamentally, cache side channels originate from the intrinsic timing difference between cache hits and misses. Attackers can use this timing difference to infer memory access patterns in contention-based cache attacks~\cite{OsvikST06,ShustermanKHMMO19} (e.g., Prime+Probe) by exploiting the limited size of cache (sets), or in shared-memory based cache attacks~\cite{YaromF14,GrussMWM16} (e.g., Flush+Reload) by manipulating and learning the cache state of a cache line shared with a victim application.
While software strives to mitigate shared-memory based cache attacks (e.g., by disabling memory deduplication and static linking of libraries) generically mitigating contention-based attacks remains difficult for software. 

Two main approaches strive to prevent contention-based attacks in hardware: Partition-based approaches~\cite{DomnitserJLAP12,WangFZMS16,WangL07} split the cache into two or more partitions and allow each partition to be used by a specific security domain only. Cache partitioning does not allow any leakage to occur between different partitions and hence provides relatively strong security, but is difficult to scale for large numbers of security domains and involves software to manage the partitions. Randomization-based approaches, on the other hand, are transparent to software and obfuscate cache side channel leakage rather than prevent it completely to allow for more efficient cache utilization.

Among the randomization-based approaches, cache-set randomization (e.g., CEASER~\cite{Qureshi18}, CEASER-S~\cite{Qureshi19}, ScatterCache~\cite{WernerUG0GM19}, and PhantomCache~\cite{TanZBR20}) has recently gained much attention. These proposals encrypt cache line addresses to randomize the mapping of addresses to cache sets and prevent attackers from inferring memory access patterns from cache-set contention. 
As attackers over time can learn the mapping from observing the cache behavior, the encryption key used by cache-set randomization needs to be regularly changed. 

While cache-set randomization is a promising direction, its security is also largely dependent on the state-of-the-art of applicable attack strategies. New approaches to efficiently learn the secret address-to-set mapping~\cite{VilaKM19, PurnalGGV21, PurnalTV21} pointed out the requirement for higher re-key frequencies that hurt performance. In addition, new analysis techniques~\cite{BourgeatDYTEY20} have highlighted the possibility to even accumulate leakage across key epochs. Consequently, there is a desire to make cache-set randomization more robust.

\austin{Cite / compare to Mirage here?}

\subsection{Contribution} 
In this work, we improve randomization-based countermeasures and present \cc to increase security and achieve practical re-key intervals. Similar to NewCache~\cite{LiuWML16}, we start with the observation that for non-partitioned caches, Fully Associative (FA) caches with random replacement achieve the best side-channel resilience. Namely, FA caches with random replacement allow every cache line address to evict any other cache line irrespective of its address and past usage. Thus, FA caches with random replacement are meant to protect against fine-grained cache-set contention attacks, but the dynamic sharing of cache resources allows for more coarse-grained cache occupancy channels~\cite{ShustermanKHMMO19}. While the side-channel properties of FA caches are desirable, they are difficult to build within typical power and area constraints. With the design of \cc, we aim to approximate the behavior of FA caches with random replacement to obtain their side-channel properties and to simultaneously keep \cc's implementation practical.

We further observe that \rsclp (\rscsp) like ScatterCache~\cite{WernerUG0GM19} and CEASER-S~\cite{Qureshi19} use multiple address-to-set mappings, which effectively increase the likelihood of two addresses  contending in the cache. To approximate a FA cache with random replacement, \cc thus builds upon \rscsp and extends it with the concept of a reinserting victim cache~\cite{jouppiImprovingDirectmappedCache1990} (\vc).
The \vc in \cc decouples evictions being observed by an attacker from contention in the \rscss: a line that is evicted from the \rscss is moved to the \vc and then automatically reinserted into the \rscss using one of its alternative address-to-set mappings. As a line is moved from the \rscss to the \vc, lines may get evicted from the \vc to memory, but these evictions are unrelated to the original contention in the \rscss. Eventually, we demonstrate that \cc shows eviction patterns that are similar to FA caches with random replacement, thus aiming to prevent fine-grained cache contention attacks, at $<1\%$ performance overhead. While \cc resembles a FA cache with random replacement to enjoy its security properties, we also stress that victim caches more generally are a convenient tool to improve the security of randomized caches as they help add noise to the attacker's observations without performance degradation. 


The paper is organized as follows. Section~\ref{sec:background} gives background about cache attacks and countermeasures. We present \cc in Section~\ref{sec:cc} and evaluates its security and performance in Section~\ref{sec:security} and Section~\ref{sec:evaluation}, respectively. We generalize the victim cache idea in Section~\ref{sec:variants}, compare with related work in Section~\ref{sec:comparison}, and finally conclude in Section~\ref{sec:conclusion}.

\section{Background}
\label{sec:background}

In this Section, we present background on cache attacks and state-of-the-art countermeasures. 

\subsection{Cache Attacks}

Modern computing systems make extensive use of caches to bridge the performance gap between the CPU and memory. However, cache structures have also been shown to allow for information leakage in side-channel attacks. These cache side-channel attacks make use of the intrinsic timing difference observed depending on whether a memory request hits or misses in the cache. An attacker can use this information about cache hits and cache misses to understand whether a victim application has accessed a memory location. Cache side-channel attacks thus reveal memory access patterns that can be used to infer sensitive information such as user behavior~\cite{ShustermanKHMMO19} and cryptographic keys~\cite{LiuYGHL15}.

There are two main categories of cache side channels: (1) Contention-based channels make use of contention in the shared cache resource, which reveals information about the cache usage of other applications. E.g., Prime+Probe~\cite{OsvikST06} uses fine-grained cache-set contention in set-associative caches to infer memory access patterns at a high frequency, and cache occupancy~\cite{ShustermanKHMMO19} attacks analyze coarse-grained contention that reveals how much cache other applications use. (2) Shared-memory based channels make use of cache lines being shared between two applications (e.g., via shared libraries or memory deduplication) and allows attackers to accurately determine whether a specific cache line has been accessed or not, e.g., via Flush+Reload~\cite{YaromF14}. 

\subsection{Cache Attack Countermeasures}

\subsubsection{Software}

While software can in principle mitigate shared-memory based channels by simple avoidance of shared memory (e.g., through disabling memory deduplication and static linking of libraries) contention-based channels are harder to mitigate from within software. For instance, in cache coloring~\cite{ZhangDS09} the operating system (OS) adjusts the virtual-to-physical mapping to map different portions of the cache to different security domains. While cache coloring achieves strong isolation similar to hardware-based cache partitioning, it is difficult to manage, hard to scale for many security domains, and has the undesirable side effect of the binding memory allocation of security domains to their cache allocations. As a result, hardware-based countermeasures seem preferable to mitigate contention-based channels.

\subsubsection{Hardware}
Hardware countermeasures can coarsely be categorized in partition-based and randomization-based countermeasures. 

\paragraph{Cache Partitioning}
Cache partitioning splits the cache into multiple partitions where each partition can be used by its assigned security domains only. For instance, non-monopolizable caches~\cite{DomnitserJLAP12} constructs its partitions by assigning each security domain a distinct subset of its cache ways. An alternative approach to partitioning is cache line pinning~\cite{WangL07}, which provides a software interface to pin specific cache lines in the cache that other security domains can no longer evict. However, while way-based partitioning is difficult to scale for many partitions, cache line pinning requires individual software support and its extensive use can deprive other applications from cache resources.

\paragraph{Cache Randomization}
While cache partitioning aims to completely stop side-channel leakage, randomization-based approaches allow side-channel leakage between different security domains to occur and obfuscate the side-channel signal to make its exploitation sufficiently hard. An example of an ideal design is a fully associative cache with random replacement, which, in absence of shared cache lines, can only leak overall cache utilization. While the power demand for large fully associative caches is typically too high, NewCache~\cite{LiuWML16} presents a more efficient implementation variant of a fully-associative cache that uses a two-step lookup process to trade off the properties of a fully-associative design with power and implementation cost. Building upon the state-of-the-art set-associative cache, RPCache~\cite{WangL07} performs an indirect cache lookup via random permutation tables in order to randomize the address-to-cache mapping. While hiding the mapping between addresses and cache sets does not stop contention-based channels, it helps to prevent attackers from directly inferring fine-grained memory access patterns. However, implementing RPcache is challenging as it requires software to manage randomization tables for the indirect set lookup. 

A different approach is cache-set randomization: CEASER~\cite{Qureshi18} realizes efficient randomization by encrypting physical addresses before extracting the cache-set index and accessing standard set-associative caches. In addition, CEASER needs to regularly change the key to prevent attackers from learning address-to-set mappings and ensure long-term side channel resistance. The successor designs CEASER-S~\cite{Qureshi19} and ScatterCache~\cite{WernerUG0GM19} introduce \rsclp (\rscsp) that improve security by skewing~\cite{seznecCaseTwowaySkewedassociative1993} the cache by its divisions. Each division consists of a few cache ways and \rscsp derive a different set index for each division by using a different encryption key. \rscsp make it hard for attackers to find a minimal set of addresses that map to exactly the same cache locations as a victim address of interest (i.e., an eviction set), because the probability that two addresses collide in all cache divisions is very low, i.e., $p = s^{-d}$ for $s$ cache sets and $d$ cache divisions. To overcome this low probability, attackers can use more likely partial collisions in the \rscss. Such partially conflicting addresses collide with the victim, e.g., in a single division only, but also have smaller probability to evict the victim address (or observe a victim access), i.e., $d^{-2}$ if an address collides with the victim address in a single division. An alternative improvement over CEASER to \rscsp is PhantomCache~\cite{TanZBR20}, which computes several randomized cache-set indices to look up multiple cache sets in parallel. 

Previous designs based on cache-set randomization and \rscsp have been shown susceptible to advanced attack strategies~\cite{PurnalTV21}, such as the Group Elimination Method~\cite{VilaKM19,Qureshi19} and \ppp~\cite{PurnalGGV21}. For instance, \ppp (1) \textsf{primes} the cache with a set of candidate attacker addresses, (2) removes candidate addresses that miss in the cache (\textsf{prune}), (3) triggers the victim to access the address of interest, and (4) \textsf{probes} the remaining set of candidate addresses for cache hits / misses. A candidate address missing in the cache has a conflict with the victim in at least one division. While advanced attack strategies such as \ppp do not break the \rscsp' security entirely, they demand for re-keying rates that are significantly higher than originally envisioned. Ultimately, these higher re-keying rates can significantly degrade performance and render the overall design impractical. While PhantomCache appears more resilient to these strategies~\cite{arxivCacheFX2022}, the design features high associativity and thus higher power consumption. 

Mirage~\cite{saileshwar2021mirage} recognizes the security benefits of fully-associative caches and builds upon the ideas of \rscsp to make fully-associative caches more practical to implement. Mirage introduces a level of indirection between a skewed, randomized, over-provisioned tag array and a global data array and allows for relocating tag entries to make sure replacement decisions are made globally and evictions in the tag array become rare. However, the over-provisioning of the tag array and indirection also leads to storage/area requirements 20\% over the baseline set-associative cache, which may be prohibitive for some applications.


\section{\cc}
\label{sec:cc}

In the following we describe \cc, a new randomized cache design to increase the security against contention-based cache attacks and thus reduce re-keying rates.

\subsection{Idea}

Caches in modern computing systems leak memory access patterns between several parties sharing the same cache. Fundamentally, this address leakage stems from two sources: First, the organization of caches in cache sets and cache ways uses a deterministic mapping from addresses to sets. This mapping allows to conclude about address information from contention in cache sets. Second, replacement policies like LRU reveal information about the access order and timing.

On the contrary, Fully Associative (FA) caches with random replacement do not leak address information as the selection of physical cache lines for insertion and eviction is entirely random. This effectively reduces the leakage for a cache that is shared between distrusting users. Contrary to partitioned cache designs, such FA caches with random replacement inevitably leak cache occupancy due to sharing the cache resource, but can attain better resource utilization. \thomas{some experiments / reference emphasizing that would be great}

While FA caches exhibit desirable security properties, their area and power demand are prohibitive for larger caches. To improve on these implementation aspects, this work presents \cc. \cc mimics the statistical properties of FA caches to inherit their security properties at lower implementation cost.

\subsection{Concept}

\begin{figure*}[t]
 \begin{center}
 \includegraphics[width=1.4\columnwidth]{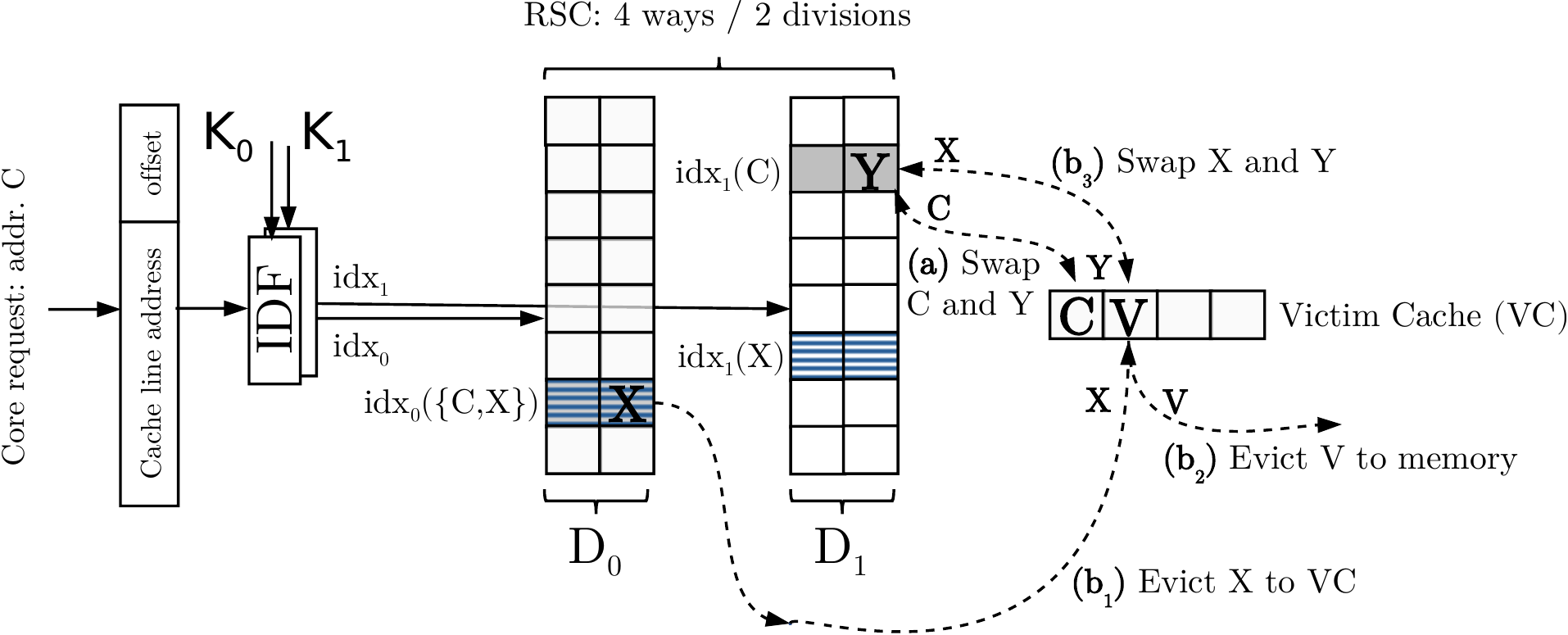}
\end{center}
\caption{Concept of Chameleon Cache for handling a core request $C$. Request $C$ and address $X$ translate to the blue striped and the gray shaded lines, respectively. Flows \textbf{($\mathbf{a}$)} and \textbf{($\mathbf{b}$)} show the cases for a hit in the \vc and eviction from the \rscss, respectively. }
\label{fig:cc_concept}
\end{figure*}

Prior work on \rscsp has demonstrated a significant security improvement over set-associative caches by making it very unlikely that two addresses map to the same set of cache lines. However, observation of eviction patterns using techniques like Prime+Prune+Probe\cite{PurnalGGV21} still allows to learn about contention between addresses. To overcome this issue, conceptually, \cc combines \rscsp with a small, fully associative victim cache (\vc) that automatically reinserts elements that have been evicted from the \rscss to the \vc. This \vc breaks the link between evictions from \cc and contention in the \rscss, results in eviction patterns similar to a FA cache with random replacement, and thus makes it harder for attackers to successfully learn \rscss contention.



The example in \autoref{fig:cc_concept} depicts the concept in more detail: When the core issues a request to address $C$, \cc first computes the indices the request maps to in each \rscss division via an index derivation function (IDF) and performs look-ups to both the \rscss and the \vc in parallel. 
\begin{enumerate}
\item If the request hits in the \rscss, the line is simply returned. 
\item If the request hits in the \vc, the line is returned and also reinserted in one of the \rscss sets previously determined for line $C$. If this reinsertion of $C$ conflicts with another line $Y$ in the \rscss, \textbf{($\mathbf{a}$)} $Y$ is put in place of $C$ in the \vc, i.e., $C$ and $Y$ are swapped.
\item If the request to address $C$ misses in both the \rscss and the \vc, the line $C$ is fetched from memory and inserted into one randomly chosen \rscss division, where the concrete set index has been determined for line $C$ before. \newline
Upon insertion of line $C$ into the \rscss, division $D_0$ in \autoref{fig:cc_concept}, \textbf{($\mathbf{b_1}$)} this line $C$ may conflict with some line $X$ stored in the \rscss before, which results in the line $X$ to be moved to the \vc.
Moving the line $X$ to the \vc may cause \textbf{($\mathbf{b_2}$)} eviction of a line $V$ previously present in the \vc to the memory.
\end{enumerate}
 
Later, the \vc tries to reinsert $X$ into the \rscss, likely in a different cache way or division. If this reinsertion of $X$ conflicts with another line $Y$ in the \rscss, \textbf{($\mathbf{b_3}$)} $Y$ is put in place of $X$ in the \vc, i.e., the lines $X$ and $Y$ are swapped.

\subsection{Specification}

\cc uses an \rscss with $s$ sets, $w$ ways and $N = s \cdot w$ lines that are organized in $1 \leq d \leq w$ divisions. The divisions $D_0,D_1,\dots,D_{d-1}$ each consist of $\frac{w}{d}$ ways, where each \rscss way is mapped to a single division only. The \rscss is skewed by its divisions: when accessing the cache, a different index $idx_i$ is used to select the set $S_{i,idx_i}$ in each division $D_i$. The \rscss uses an IDF to compute the divisions' set indices $idx_0, idx_1,\dots,idx_{d-1}$ from the requested cache line address. The pseudo-random mapping given by the IDF is regularly changed, e.g., by changing its keys. 

\cc uses a fully associative victim cache (VC) with $w_{VC}$ ways. Cache lines evicted from the \rscss are moved to the \vc, which performs automatic reinsertion into the \rscss to increase security and the interval for changing the IDF mapping. The \vc maintains two indices $idx_{VC,insert}$ and $idx_{VS,reinsert}$ to keep track of the last item that has been inserted into the \vc and re-inserted into the \rscss, respectively. In addition, requests that hit in the \vc result in automatic reinsertion of the respective line into the \rscss.

\cc uses a set of different algorithms to perform its operations. \textbf{Init} in \autoref{alg:init} initializes the cache. \textbf{IDF} in \autoref{alg:idf} performs the mapping of cache line addresses to \rscss sets. \textbf{Lookup} in \autoref{alg:lookup} describes the lookup of a cache line address in \cc. \textbf{RSC Insert} in \autoref{alg:rsc_insert} specifies the insertion of a new cache line into \cc. \textbf{RSC Reinsert} in \autoref{alg:rsc_reinsert} performs reinsertion of a cache line from the \vc to the \rscss. \textbf{Automatic RSC Reinsert} in \autoref{alg:automatic_reinsert} is periodically triggered to initiate automatic reinsertion of lines from the \vc to the \rscss. Note that the respective counters $idx_{VC,insert}$ and $idx_{VC,reinsert}$ automatically wrap around when they reach $w_{VC}$. For simplicity, all algorithms omit the wrap-around logic.

\begin{algorithm}[t]
\textbf{Input:} \rscss with $w$ ways, $s$ set indices and $d$ divisions, \vc \\
\textbf{Output:} Initialized \rscss and \vc 
\begin{algorithmic}[1]
\For{$0\leq i < d, 0 \leq j < s, 0 \leq k < \frac{w}{d}$} 
\State $RSC[i][j][k] \gets \bot$ 
\EndFor
\State $VC[i] \gets \bot \;\; \forall \;\; 0 \leq i < w_{VC}$ 
\State $idx_{VC,insert} \gets 0$
\State $idx_{VC,reinsert} \gets 0$
\end{algorithmic}
\caption{Init}
\label{alg:init}
\end{algorithm}

\begin{algorithm}[t]
\textbf{Input:} address $A$, keys $K_0,...,K_{d-1}$ \\
\textbf{Output:} indices $idx_0,...,idx_{d-1}$ for $d$ \rscss divisions
\begin{algorithmic}[1]
\For{$0 \leq i < d$}
 \State $A_{enc,i} \gets E_{K_{i}}(A)$
 \State $idx_{i} \gets \lceil A_{enc,i} \rceil^{\log_2{s}} $ \;\;\;\; // Slice out $\log_2{s}$ bits
\EndFor\\
\Return $idx_{0},...,idx_{d-1}$
\end{algorithmic}
\caption{Index Derivation Function (IDF)}
\label{alg:idf}
\end{algorithm}

\begin{algorithm}[t]
\textbf{Input:} address $A$, keys $K_0,...,K_{d-1}$ \\
\textbf{Output:} data at address $A$
\begin{algorithmic}[1]
\State $idx_{0},...,idx_{d-1} \gets IDF(A,K_0,...,K_{d-1})$
\State Hit $ \gets false$
\For{$0 \leq i < d$}
 \For{$0 \leq j < \frac{w}{d}$}
    \If{$RSC[i][idx_{i}][j].tag = A$} 
      \State Data $\gets RSC[i][idx_{i}][j].data$
      \State Update $RSC[i][idx_{i}].lru\_state$ if required
      \State Hit $\gets true$
   \EndIf
 \EndFor
\EndFor
\For{$0 \leq i < w_{VC}$}
  \If{$VC[i].tag = A$}
    \State Data $\gets VC[i].data$
    \State Hit $\gets true$
    \State $RSCReinsert(i)$
  \EndIf
\EndFor
\If{$\overline{\textnormal{VCHit}}$ \textbf{and} $\overline{\textnormal{RSCHit}}$}
  \State Data $\gets memory[A]$
  \State $RSCInsert(Data, idx_0, ..., idx_{d-1})$
\EndIf\\
\Return Data
\end{algorithmic}
\caption{Lookup}
\label{alg:lookup}
\end{algorithm}

\begin{algorithm}[t]
\textbf{Input:} data $D$, address $A$, indices $idx_0,...,idx_{d-1}$ \\
\begin{algorithmic}[1]
\State $\hat{d} \overset{\$}{\gets} \{0,...,d-1\}$
\State In set $RSC[\hat{d}][idx_{\hat{d}}]$: select victim line index $v$ according to replacement policy
\If{$RSC[\hat{d}][idx_{\hat{d}}][v].tag \neq \bot$}
  \State $idx_{VC,insert} \gets idx_{VC,insert} + 1$
  \If{$VC[idx_{VC,insert}].tag \neq \bot$}
    \State Evict line at $VC[idx_{VC,insert}]$ to memory
  \EndIf
  \State $VC[idx_{VC,insert}] \gets RSC[\hat{d}][idx_{\hat{d}}][v]$
\EndIf
\State $RSC[\hat{d}][idx_{\hat{d}}][v].data \gets D$
\State $RSC[\hat{d}][idx_{\hat{d}}][v].tag \gets A$
\State Update replacement bits in set $RSC[\hat{d}][idx_{\hat{d}}]$ if necessary
\end{algorithmic}
\caption{RSC Insert}
\label{alg:rsc_insert}
\end{algorithm}

\begin{algorithm}[t]
\textbf{Input:} VC line index $idx_{vc}$ to be reinserted into the RSC\\
\begin{algorithmic}[1]
\State $idx_0,...,idx_{d-1} \gets IDF(VC[idx_{vc}].tag)$
\State $\hat{d} \overset{\$}{\gets} \{0,...,d-1\}$
\State In set $RSC[\hat{d}][idx_{\hat{d}}]$: select victim line index $v$ according to replacement policy
\State Swap $RSC[\hat{d}][idx_{\hat{d}}][v]$ and $VC[idx_{vc}]$
\State Update replacement bits in $RSC[\hat{d}][idx_{\hat{d}}]$ if necessary
\end{algorithmic}
\caption{RSC Reinsert}
\label{alg:rsc_reinsert}
\end{algorithm}

\begin{algorithm}[t]
\begin{algorithmic}[1]
\While{$idx_{VC,reinsert} < idx_{VC,insert}$}
  \State $RSCReinsert(idx_{VC,reinsert})$
  \State $idx_{VC,reinsert} \gets idx_{VC,reinsert} + 1$
\EndWhile
\end{algorithmic}
\caption{Automatic RSC Reinsert}
\label{alg:automatic_reinsert}
\end{algorithm}

\paragraph{Index Derivation Function}

\autoref{alg:idf} implements the IDF using a cryptographic block cipher $E$. However, the IDF may also be implemented based on other primitives, such as (keyed) hash functions $H$. A suitable IDF must guarantee that (1) the keys remain secret as attackers observe addresses mapping to the same index (\textit{collisions}), (2) addresses that have index collisions under one key $K_A$ have an index collision with a different key $K_B$ only with negligible probability. Moreover, the IDF should be efficient to implement to ensure low access latencies. 

\thomas{We may talk about re-keying intervals, but that is better to be done based on security analysis results}

\thomas{we may argue about performance due to higher associativity}

\subsection{Indistinguishability}

A main security requirement for \cc is the \textit{indistinguishability} of \rscss and \vc. Namely, implementations of \cc must ensure that attackers cannot distinguish whether a line is in the \rscss or in the \vc. Otherwise, attackers would be able to recognize contention in the \rscss by monitoring when lines are (temporarily) moved to the \vc. The requirement of indistinguishability implies that implementations must make sure that (a) \rscss and \vc show the same access latency for cache hits and (b) there are no observable side effects when lines transition between the \rscss and \vc and vice versa. Thus, a first step to achieve indistinguishability is a cache pipeline that returns the results from \vc and \rscss in the same pipeline stage as this can provide the same access latency for both \rscss and \vc. However, this list is non-exhaustive and a concrete implementation may require additional measures to be taken to guarantee indistinguishability.
 
\section{Security Analysis} \label{sec:security}

As \autoref{sec:cc} showed, \cc extends \rscsp with a \vc and automatic reinsertion to improve the security and thus reduce required re-keying intervals. 
In the following, we analyze the security of \cc and compare it to other works. Our analysis consists of a qualitative analysis of the cache's eviction behavior, a probabilistic analysis that formalizes the relative difficulty of contention-based attacks with \cc, and a quantitative empirical analysis using a cache attack simulation framework.
While our specification of \cc is agnostic to the replacement strategy of the \rscss, we note that stateful replacement strategies like LRU come with additional side-channel leakage~\cite{xiongLeakingInformationCache2020} and hence we focus our analysis on random replacement.

\subsection{Victim Cache}
\label{sec:sec_vc}

\cc extends \rscsp with a \vc to break the direct link between cache conflicts in the \rscss and the cache misses that may be observed, e.g., in \ppp. As \autoref{fig:cc_concept} shows, when a cache line $C$ is inserted into the \rscss and evicts another line $X$ to the \vc, the cache conflict is hidden as both lines $C$ and $X$ will hit in the cache afterwards. In addition, a line $V$ potentially being evicted from the \vc to the memory is uncorrelated to the cache conflict in the \rscss. 

Without reinsertion of the evicted line, however, the attacker may be able to inspect the contents of the \vc and yet learn about the conflict. For instance, an attacker could access a set of random, uncached addresses to force cache lines being moved to the \vc. This would eventually flush the lines previously stored in the \vc to the memory, giving the attacker a measurable side-channel about the \vc. Note however that profiling \rscss cache contention via flushing the \vc adds noise to the attacker's measurements, i.e., the attacker will observe cache misses on lines that were previously present in the \vc but are unrelated to the \rscss contention introduced by accessing $C$. Intuitively, the number of false positives grows with the size of the \vc. 

To prevent attackers from learning about \rscss contention via the \vc, \cc automatically reinserts cache lines, which have been evicted from the \rscss to the \vc, back into the \rscss (cf. \autoref{alg:automatic_reinsert}). As a cache line $X$, which has been evicted by line $C$ from the \rscss to the \vc, is reinserted into the \rscss, two possible situations can arise:
\begin{enumerate}
 \item \textbf{Reinsertion into a different cache way}:
 Reinsertion of a line $X$ to the \rscss results in another line $Y$ being placed into the \vc, i.e., $X$ and $Y$ are swapped. This makes the \rscss conflict between $C$ and $X$ invisible, as eventually both will be stored in the \rscss again. Cache line $Y$ in the \vc, on the other hand, is either invalid or does not directly relate to the contention between $C$ and $X$ in the \rscss. 
 
 \item \textbf{Reinsertion into the same cache way}: 
 $X$ is reinserted into the same cache way it was evicted from, i.e., $X$ and $C$ are swapped. As a result, $X$ is stored in the \rscss and $C$ in the \vc, thus making the previous \rscss contention invisible. 
 
 Note that the attacker may be able to re-access $C$ in order to bring $C$ back into the \rscss. In this case, a line $Z$ is swapped with $C$, which results in $Z$ and $C$ being in the \vc and \rscss, respectively. This line $Z$, which may be the original line $X$, directly contends with $C$ in the \rscss. While an attacker might be able to learn $Z$ by flushing the \vc, using $Z$ meaningfully remains hard. Namely, whenever $Z$ evicts $C$ from the \rscss, the automatic reinsertion mechanism will move $C$ back into the \rscss and hence make the conflict invisible. 
\end{enumerate}

\subsection{Second-Order Collisions}

The principle of automatic reinsertion still affects the cache state and intuitively bears second-order leakage. More concretely, and as \autoref{fig:cc_concept} shows, accessing a line $C$ that conflicts with a line $X$ can lead to another line $Y$ ending up in the \vc, potentially making this line $Y$ visible to attackers who are able to flush the \vc. While $Y$ is not directly conflicting with the line $C$, $C$ and $Y$ are connected via the conflicting line $X$ that goes to the \vc and back to the \rscss. $X$ conflicts both with $C$ and $Y$ in at least one division each and may thus serve as a proxy in a cache attack. Note, however, since there are multiple divisions in the \rscss, this does not imply that $C$ and $Y$ conflict. We in the following denote such addresses $Y$ \soaddresses.

Attackers may be able to learn about \soaddresses and use them in an attack to measure \rscss contention with $C$. In \autoref{fig:cc_concept}, $C$ may evict line $Y$ to the \vc via the proxy $X$ and flushing $Y$ from the \vc as described in \autoref{sec:sec_vc} may then allow an attacker to observe the contention. Yet, \soaddresses are hard to exploit in practice for multiple reasons. 
\thomas{a figure may be helpful to show this type of attack}
\begin{enumerate}
 \item \textbf{Indistinguishability: } For profiling strategies like \ppp, an attacker observing a cache miss after flushing the \vc cannot determine whether they sampled a \soaddress or a completely unrelated address. This effectively increases the number of addresses needed for an eviction set and hence noise.
 \item \textbf{Unknown proxy address: } Attackers do not know the proxy address $X$, because it cannot be observed. However, the \soaddress $Y$ is only valuable for attackers if they know and insert $X$ before, i.e., $X$ is a proxy for $Y$ and is required to evict $C$. 
 
 Moreover, since $X$ is unknown, a \soaddress $Y$ that is collected by the attacker is as good as a random address: Without $X$, $Y$ has the same probability as a randomly chosen address to evict $C$.  
 \item \textbf{Prevalence of proxy addresses: } An attacker observing a miss on the \soaddress $Y$ via proxy $X$ is unlikely to find another proxy $X'$ which collides with both $C$ and $Y$ in different divisions. A randomly chosen address is a proxy for $C$ and $Y$ with a probability $p \approx \frac{w^2}{s^2}$ if $d=w$, e.g., 1 in 16384 addresses are a suitable proxy in a cache with 2048 sets and 16 ways. However, an arbitrarily chosen address itself has higher probability of directly featuring a partial collision with $C$, roughly $p = 1 - \frac{(s-1)^w}{s^w} \approx 2^{-8}$ for the same cache configuration. 
 
 Note that that when $X$ contends with $Y$ and $C$ in the same division, $Y$ may directly evict $C$ from the \rscss, but automatic reinsertion of $C$ will make this contention invisible. 
\item \textbf{Success probability: } Even if attackers know the proxy $X$, the probability of evicting the line $C$ from the \rscss to the \vc using the \soaddress $Y$ and vice versa is low. Namely, this approach requires (1) $C$ and $X$ to reside in the correct cache divisions before, (2) $Y$ needs to be inserted such as to evict $X$, and (3) $X$ must be reinserted such as to evict $C$ to the \vc. For random replacement, \rscss eviction using \soaddresses thus has a success probability of only $w^{-4}$, e.g., $2^{-16}$ for a 16-way cache. For a 16-way cache with 2048 lines, it would hence require more addresses than fit in the cache to with high probability evict the target address into the \vc by using \soaddresses. Moreover, once $C$ has been moved to the \vc, attackers further need to flush the \vc with random addresses, which adds more noise through contention required in the \rscss.
\end{enumerate}

\subsection{Relative Eviction Entropy}

\begin{figure}
 \includegraphics[width=\columnwidth]{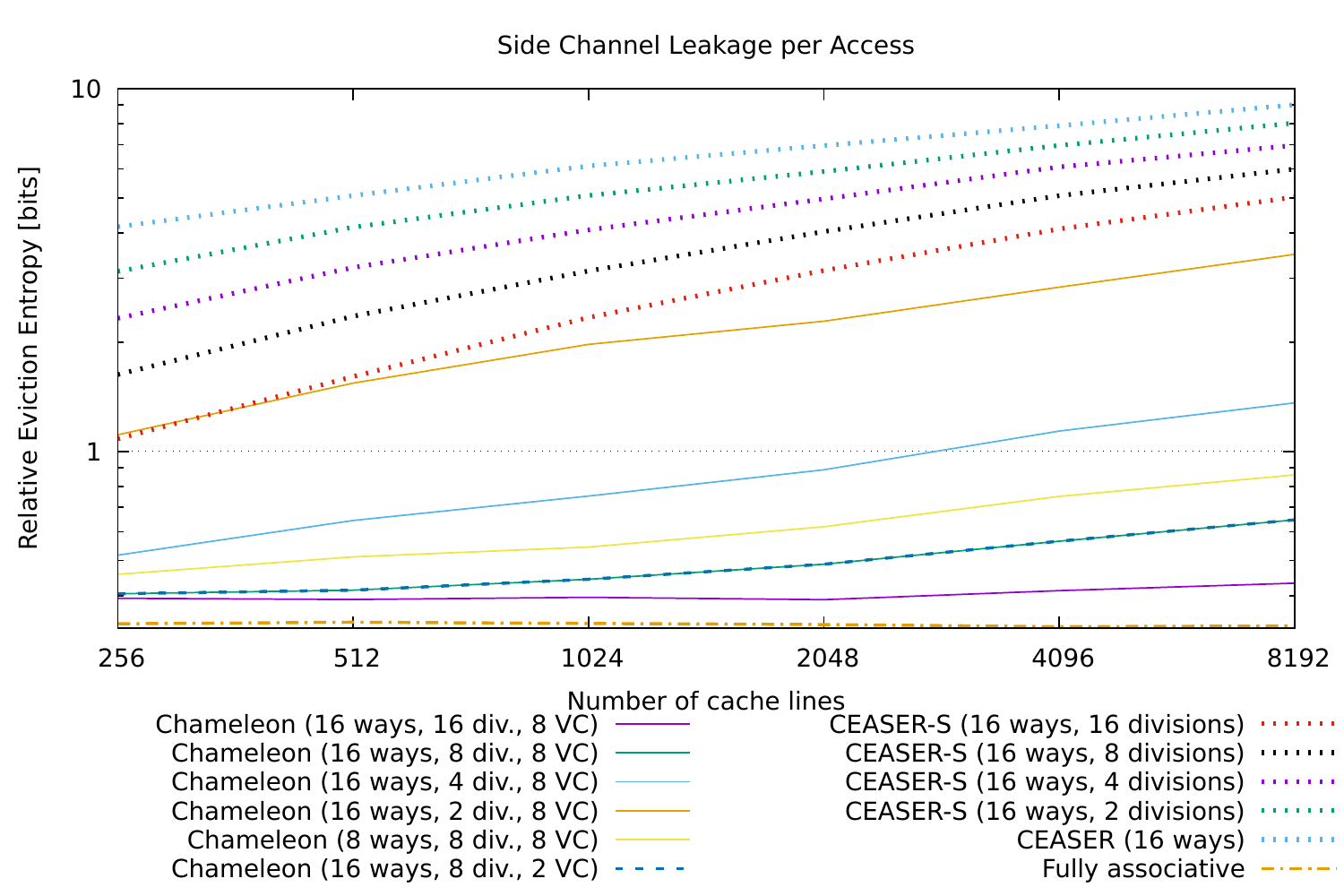}
 \caption{Relative eviction entropy for \cc compared to CEASER, CEASER-S, and a fully associative cache. }
 \label{fig:entropy}
\end{figure}

We used the cache security framework CacheFX~\cite{arxivCacheFX2022} to implement a model for \cc and comprehensively test and compare it to state-of-the-art cache designs. In \autoref{fig:entropy}, we evaluated the relative eviction entropy of \cc, CEASER, CEASER-S, and a fully associative cache for increasing cache sizes and with all these caches using random replacement. \autoref{fig:entropy} shows that the information leakage is significantly lowered for \cc compared to prior cache randomization techniques. For instance, adding a victim cache to 16-way 8192-line CEASER-S with 16 divisions reduces information leakage per eviction from 5 to 0.4\,bits. Further note that the relative eviction entropy is the same for instances of \cc that only differ in their \vc size.

\subsection{Eviction Set Success Rate}

\begin{figure}
 \includegraphics[width=\columnwidth]{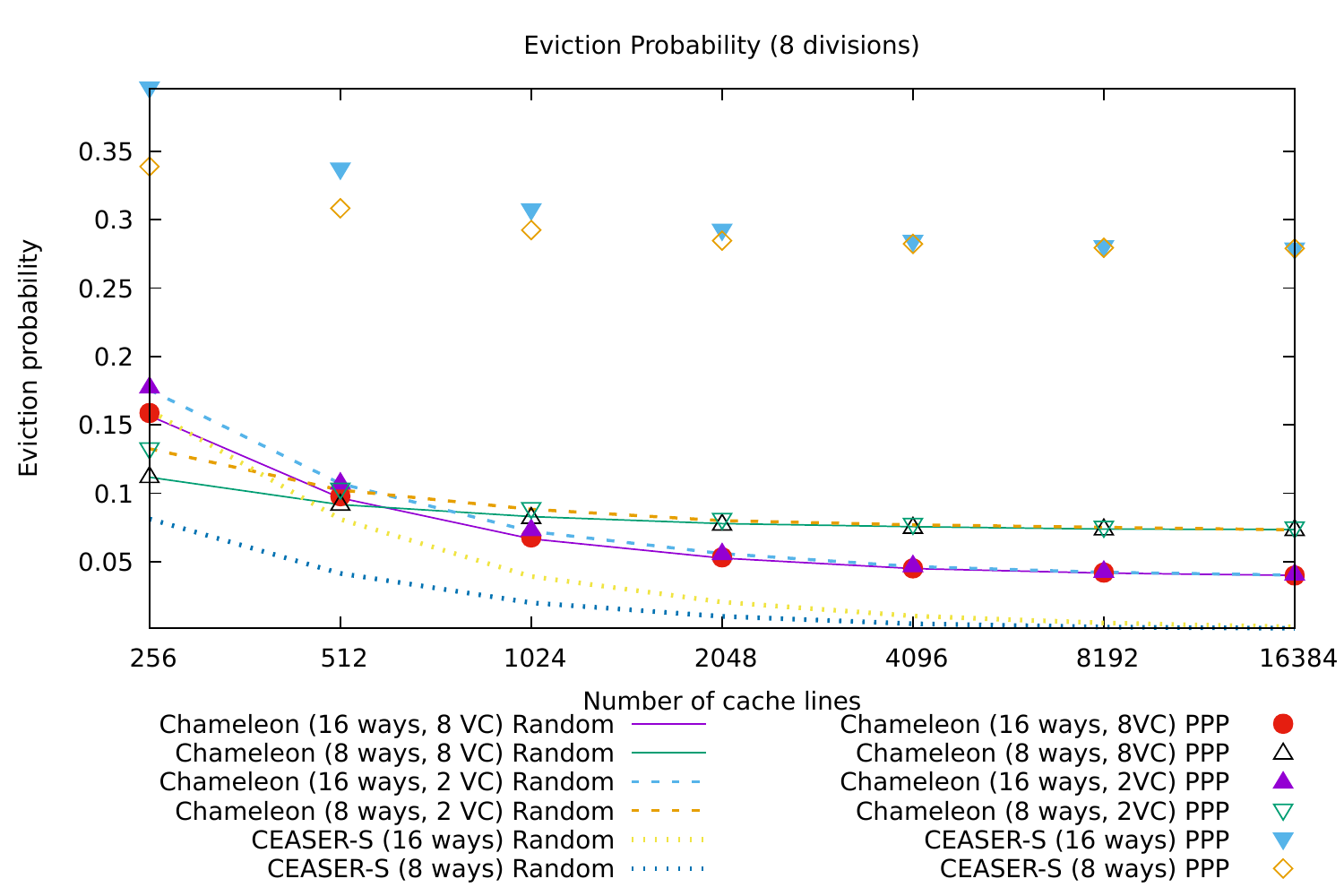}
 \caption{Eviction success rates of eviction sets constructed via \ppp compared to randomly selected addresses. All caches are operated with 8 divisions }
 \label{fig:eviction_probability}
\end{figure}

To demonstrate the security of \cc w.r.t. \ppp, we compare the eviction success rate of eviction sets constructed with \ppp on \cc to the eviction success rate for a set of randomly chosen addresses.
Using CacheFX~\cite{arxivCacheFX2022}, we run \ppp and sample a random set of addresses 
$ M = 1000 $ times to form eviction sets of $4\cdot w$ addresses for a random target and evaluate the success rate of each eviction set. The success rate of each eviction set is determined by trying to evict the target address 1000 times and compute the mean over all experiments. 
We show the mean eviction success rates for different configurations of \cc and CEASER-S in \autoref{fig:eviction_probability} and the $M$ experiments. In this evaluation, we operate all caches with 8 divisions and experimented with 2 and 8 victim cache lines for \cc as well as 8 and 16 cache ways. While CEASER-S shows a strong difference in the eviction success rate for eviction sets constructed via random sampling and \ppp, \cc does not yield an observable difference. 

To investigate the properties of \cc further, we also determine the statistical variance of the $M$ eviction success rates and compute the t-value~\cite{SchneiderM15}. We reject the hypothesis that the mean success rates for eviction sets from \ppp and random sampling are equal with a confidence of 99.999\% if $|t| > 4.5$.

\begin{figure}
 \includegraphics[width=\columnwidth]{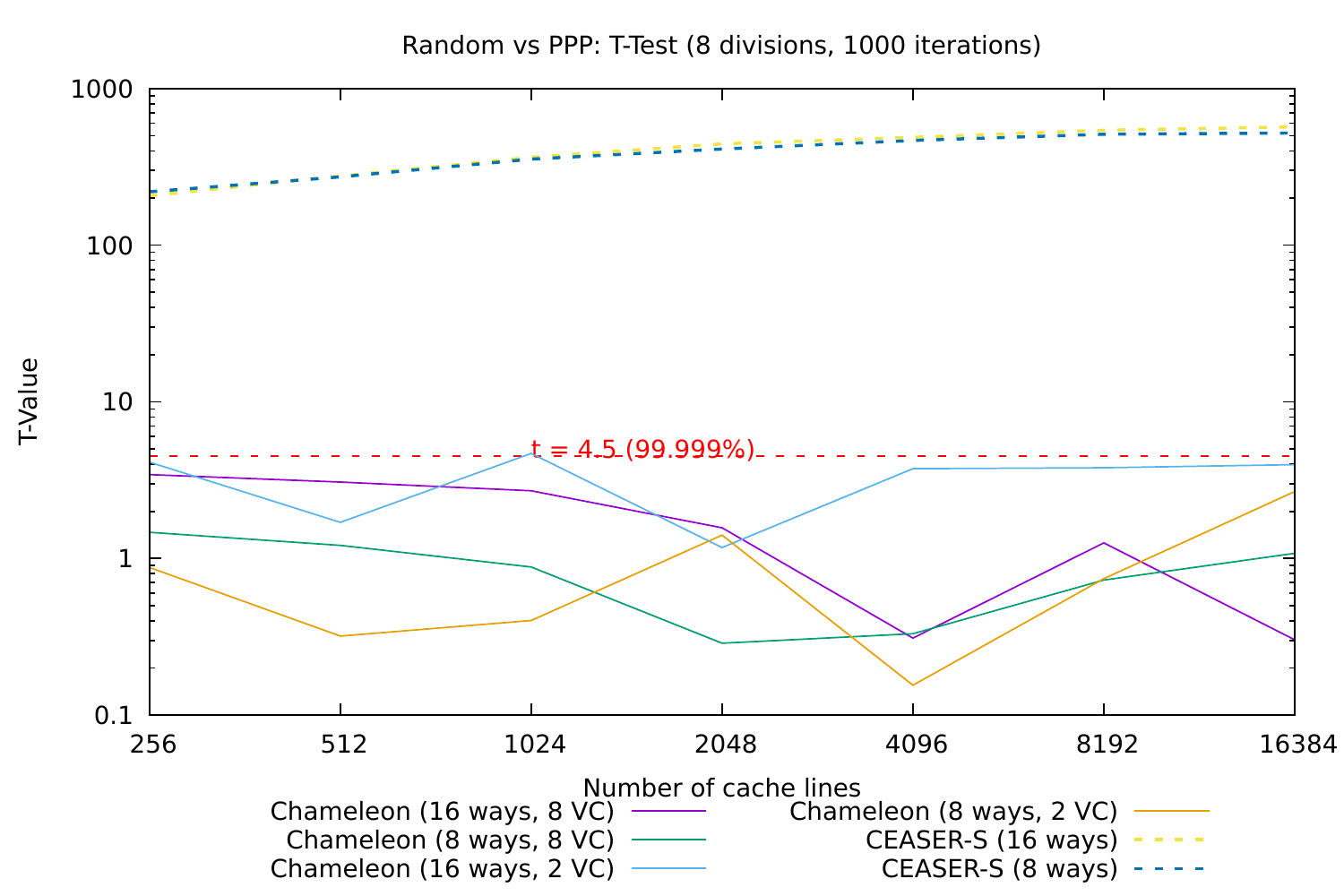}
 \caption{T-values for the eviction success rates of eviction sets constructed via \ppp compared to randomly selected addresses.}
 \label{fig:ttest_eviction_probability}
\end{figure} 

\autoref{fig:ttest_eviction_probability} shows that the t-value stays largely below the threshold of 4.5 for \cc, suggesting that attackers do not have a clear advantage from constructing eviction sets with \ppp over random sampling. On the other hand, eviction sets built with \ppp for CEASER-S have success rates clearly distinguishable to success rates of randomly assembled eviction sets. Note however that for \cc and a larger number of different experiments $M$ there is statistically measurable difference between eviction sets constructed via \ppp and random sampling according to the t-statistics. Yet, we argue that this difference is small enough not to be relevant in practice. 

\subsection{Eviction Set Profiling}

\begin{figure}
 \includegraphics[width=\columnwidth]{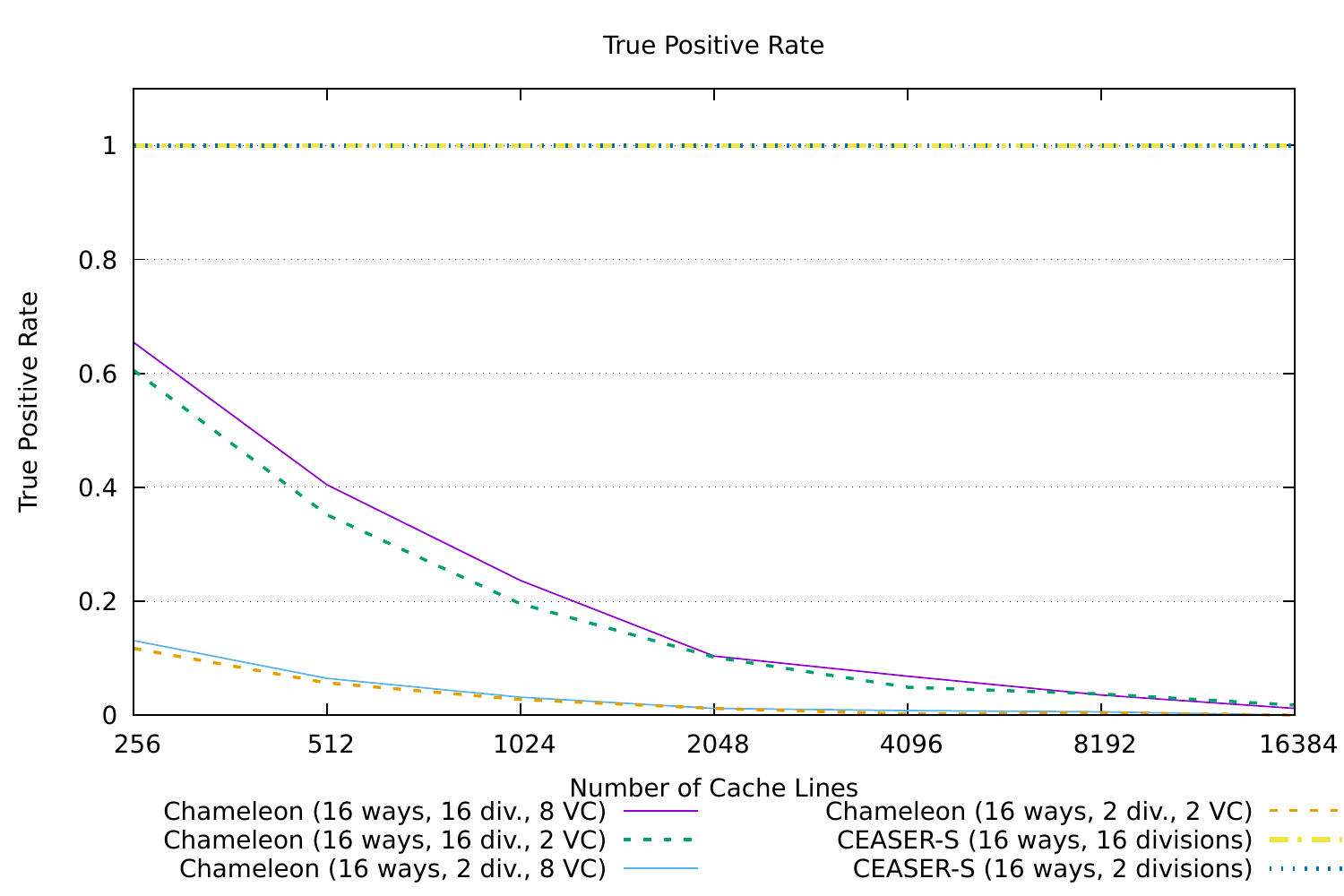}
 \caption{Rate of addresses in the eviction set that are truly conflicting with the victim address.}
 \label{fig:ppp_true_positive_rate}
\end{figure}

We evaluated the properties of \ppp, the currently most-efficient algorithm to construct eviction sets for skewed caches, in more detail. \autoref{fig:ppp_true_positive_rate} shows the fraction of addresses found by \ppp in a noise-free setting that are truly conflicting in at least one division with the victim address. While for CEASER-S this True Positive Rate (TPR) in the absence of noise is consistently 1, the TPR is clearly lower for all configurations of \cc and decreases with cache size. The TPR is generally smaller for 2 divisions than for 8 divisions, because more divisions increase the probability of random conflicts in any of the divisions. 

\begin{figure}
 \includegraphics[width=\columnwidth]{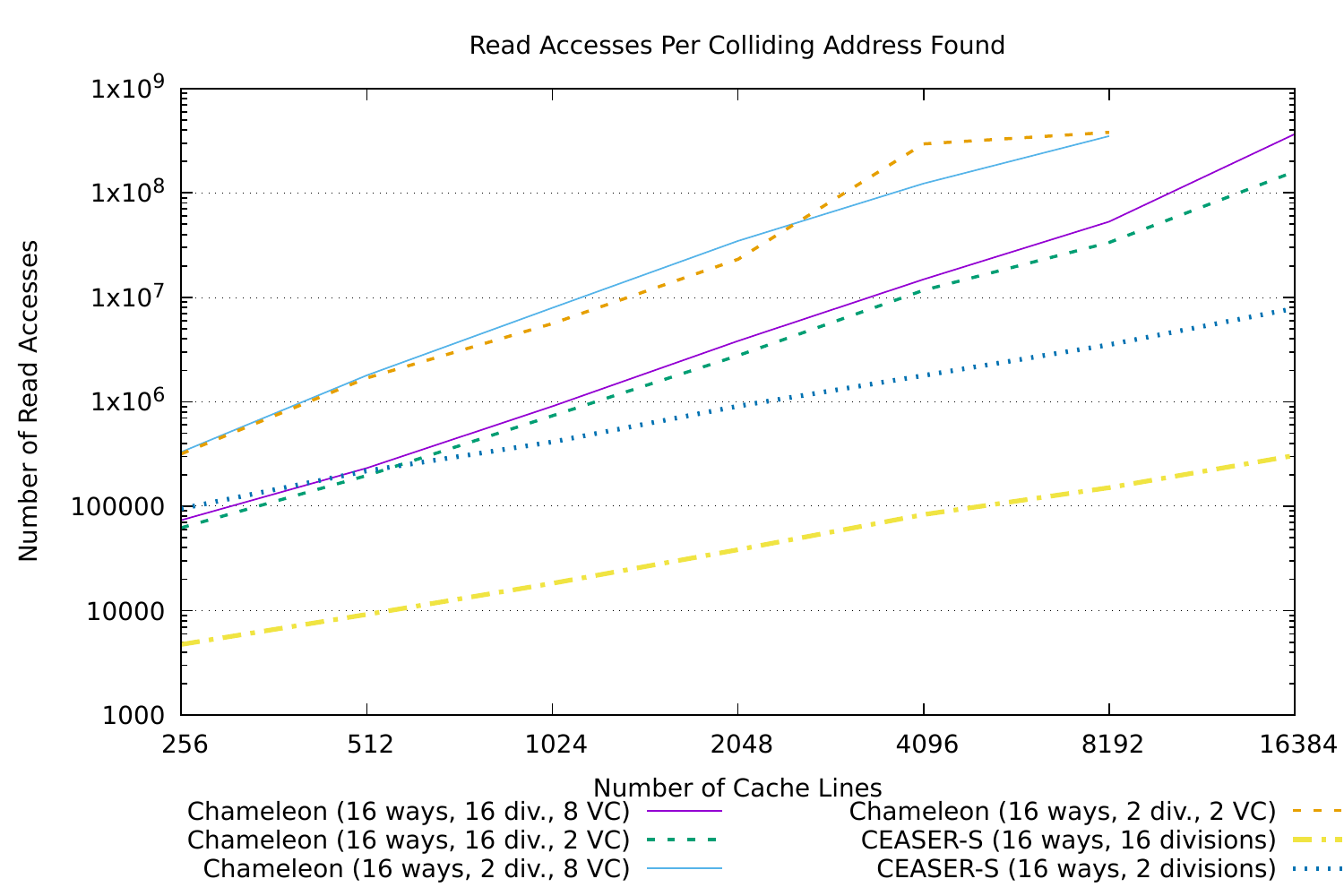}
 \caption{Number of memory accesses required by the attacker to find one truly conflicting address.}
 \label{fig:ppp_rdaccesses}
\end{figure}

\autoref{fig:ppp_rdaccesses} shows the number of read accesses that need to be done by the attacker in order to find one address that is truly conflicting with the victim address, i.e., not noise. The effort for finding such address is one order of magnitude higher for \cc compared to CEASER-S with larger instances of \cc having even higher relative profiling cost. Overall, \cc increases the cost of profiling and significantly decreases the value of the eviction sets found via \ppp.


\subsection{Cache Occupancy}

Recent work~\cite{ShustermanKHMMO19} demonstrated scenarios that exploit side-channel leakage stemming from cache occupancy. Like fully associative caches with random replacement, \cc does not protect against this cache occupancy leakage. Note however that cache occupancy leakage is very coarse-grained, has only limited temporal resolution, and is inevitable for any shared cache design. We thus believe that the security properties offered by \cc will be sufficient in most cases. 

\section{Evaluation}

\label{sec:evaluation}

\newcommand{\mnrow}[2][c]{%
      \begin{tabular}[#1]{@{}c@{}}#2\end{tabular}}
\newcommand{\ra}[1]{\renewcommand{\arraystretch}{#1}}
\begin{table}[t]
\centering
\resizebox{\linewidth}{!}{%
\ra{1.1}
\begin{tabular}{ccl}\toprule
& \RC{20}OS & \RC{20}Redhat 8 with Linux kernel 5.4.49\\
\multirow{-2}{*}{\textbf{System}} & Processor & 4 x86 OoO Cores at 3GHz\\
\midrule
& \RC{20}Predictor & \RC{20}LTAGE and Indirect Predictor, 512-entry BTB\\
& Fetch & {5 wide Fetch, Decode, Rename, 224-entry ROB} \\
& \RC{20}Dispatch & \RC{20}{8 wide Dispatch, Issue, Writeback, 97-entry IQ} \\
\multirow{-4}{*}{\rotatebox[origin=c]{90}{\textbf{Core}}} & Exec &
\begin{tabular}[c]{@{}l@{}}
4 INT ALUs, 3 INT VectU, 2 FP FMAs, \\
168/180 Phys.~Reg., 72/56-entry Ld/St~Buffer\end{tabular} \\
\midrule
& \RC{20}L1-I/D & \RC{20}\begin{tabular}[c]{@{}l@{}}32kB, 8-way,
2/4 cycles, 16-entry MSHR, Random Replacement\end{tabular} \\
& L2 & \begin{tabular}[c]{@{}l@{}}256kB, 4-way, 10 cycles,
20-entry MSHR, Random Replacement\end{tabular} \\
& \RC{20}Shared L3 & \RC{20}\begin{tabular}[c]{@{}l@{}}
16MB, 16-way, 40 cycles, 256-entry MSHR, stride prefetch \end{tabular} \\
\multirow{-5}{*}{\rotatebox[origin=c]{90}{\textbf{Memory}}} & DRAM &
\begin{tabular}[c]{@{}l@{}}8GB 4 Channel DDR4-2400, 38.4GB/s \\
Peak Bandwidth, Latency from DRAMSim2\end{tabular}\\
\bottomrule
\vspace{0.5em}
\end{tabular}
}
\caption{Gem5 Full-System Simulation Configurations~\cite{SkylakeWikiChipOnline}}
\label{tbl:gem5specs}
\end{table}

\begin{figure*}[t]
\centering
\includegraphics[width=0.8\textwidth]{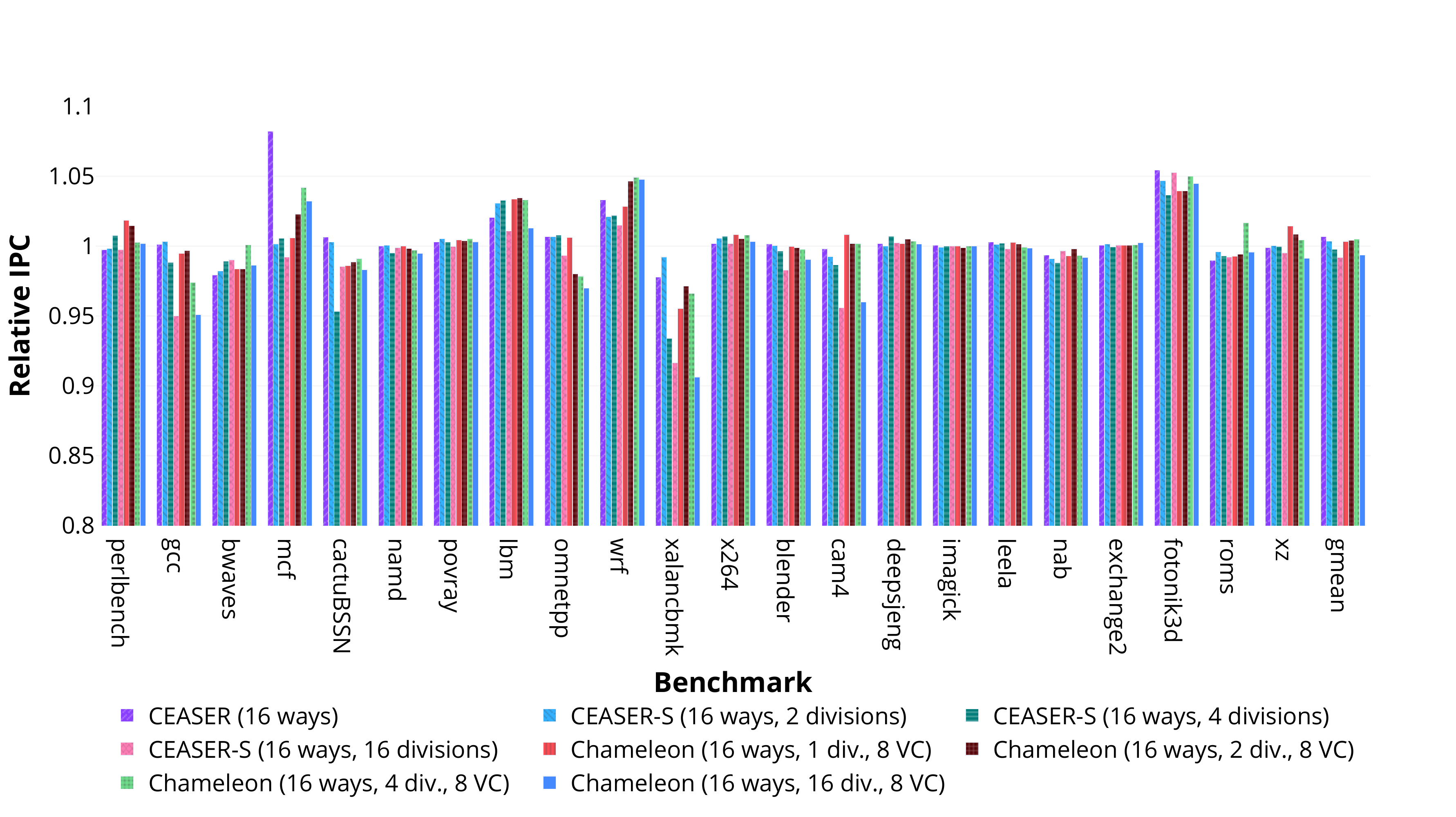}
\caption{\label{fig:specrate2017-ipc}
Relative Instructions Per Cycle (IPC) for SPECRate2017 with 4 copies.}
\end{figure*}

\begin{figure*}[t]
\centering
\includegraphics[width=0.8\textwidth]{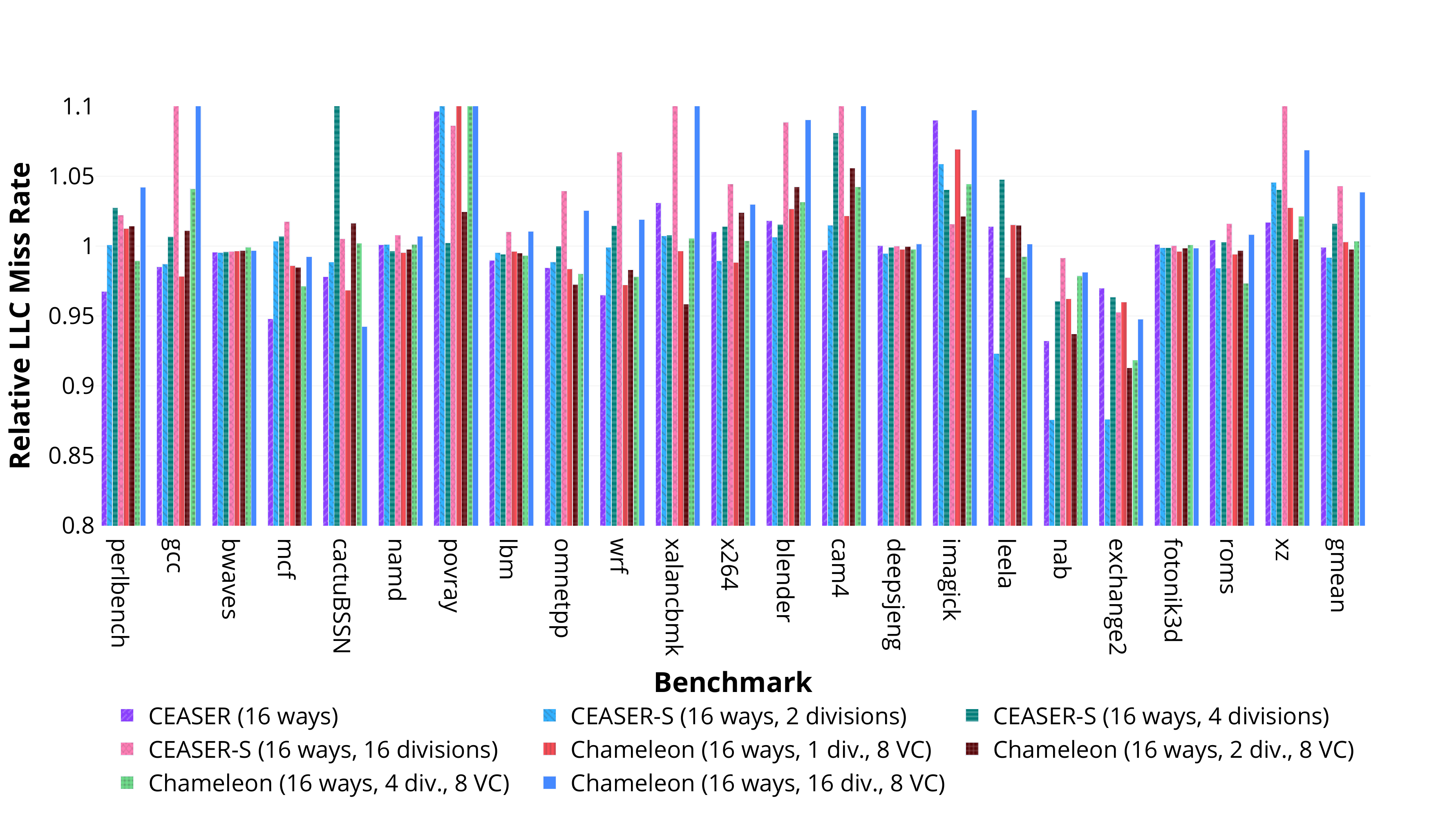}
\caption{\label{fig:specrate2017-llc-miss-rate}
Relative LLC Miss Rate for SPECRate2017 with 4 copies.}
\end{figure*}

We use the cycle-accurate gem5~\cite{binkertGem5Simulator2011, lowe-powerGem5SimulatorVersion2020} simulator to evaluate \cc. \cref{tbl:gem5specs} shows the baseline simulator configuration based on a Skylake processor. We run SPECRate2017 with 4 copies in full-system mode, and gather statistics using a sampling methodology based on pFSA~\cite{sandbergFullSpeedAhead2015} due to the long execution time of SPEC reference inputs. In this sampling methodology, we execute each benchmark using hardware virtualization (i.e., the gem5 KVM CPU) to record the total instructions and generate 1000 random samples. Next, we quickly fast-forward to each sample using KVM and then perform functional warm-up of the caches for 10 million instructions with the atomic CPU, switch to detailed warm-up for 6 million instructions, and finally record detailed statistics for 5 million instructions with the Skylake CPU.

\cref{fig:specrate2017-ipc} depicts the relative Instructions-Per-Cycle (IPC) for CEASER-S and \cc with varying division counts. Note that ScatterCache can be viewed as an instance of CEASER-S with 16 divisions. It shows that on average the relative IPC drops for CEASER-S with a higher number of divisions, whereas \cc, except  for 16 divisions, is less sensitive to the division count. This results in better average performance than for previous \rscss proposals and helps some workloads, e.g., \texttt{wrf} and \texttt{mcf}, to perform even better than the baseline design from \cref{tbl:gem5specs}. Generally, the relative performance impact is very small, i.e., $<$1\% on average, and ranges between -10\% and +5\% for individual workloads. \cref{fig:specrate2017-llc-miss-rate} further shows the relative miss rate for the shared \cc L3/LLC cache. Except for 16 divisions, \cc on average features a miss rate and relative IPC equal to the baseline. For individual workloads, miss rate and relative IPC range between -10\% and +5\%. In some cases the victim cache reinsertion can improve performance by helping spread hot sets out to other partitions.  In addition, we measured the frequency of contention events between the reinsertion of victim cache entries and incoming cache requests, but saw no conflicts in our experiments. In terms of area, \cc with an 8-way \vc must maintain 8 additional cache lines per cache slice, which amounts to $<$0.1\% additional storage for the architecture specified in \cref{tbl:gem5specs}.

\section{Variants}
\label{sec:variants}

While \cc aims to mimic a fully associative cache with random replacement and its security properties, the design principle of adding a victim cache to decouple contention in a \rscss from evictions observed by users is applicable more generally. In the following, we thus lay out several design variants that can as well increase security over a baseline \rscss.

\subsection{\cc without Reinsertion}

\begin{figure}[h]
\centering
\includegraphics[width=0.5\textwidth]{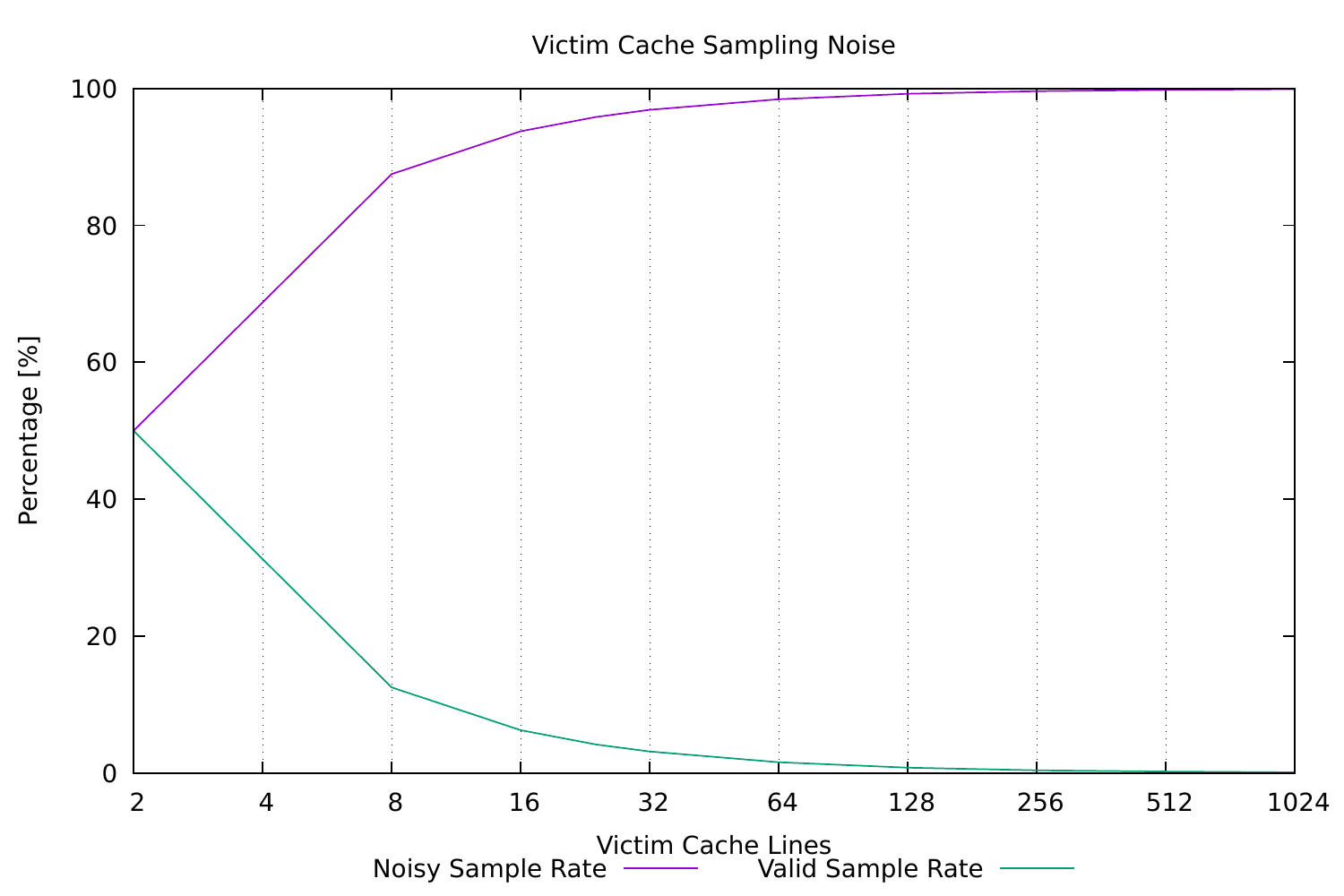}
\caption{\label{fig:vcnoise}
Rate of noisy samples and truly conflicting samples when performing eviction set profiling~\cite{PurnalTV21} in randomized set-associative caches with various victim cache sizes.}
\end{figure}

A first simplification of \cc is to omit its automatic reinsertion functionality, i.e., to simply extend a \rscss with a fully associative \vc. This fully associative victim cache can use first-in first-out replacement like \cc, or implement random replacement to add more noise to the design's eviction patterns. The disadvantage of omitting automatic reinsertion is that contention in the \rscss is more easily observable if attackers can flush the \vc, e.g., by creating contention in the \rscss. However, as indicated in \autoref{sec:sec_vc} and as depicted in \autoref{fig:vcnoise}, this adds noise  and thus profiling cost proportional to the size of the \vc and hence helps reduce re-key rates accordingly.

We expect random accesses stemming from the system's background noise to be equivalent to an attacker creating pseudo-random contention in the \rscss in terms of their ability to flush the \vc. Consequently, some open questions are (1) to what extent system noise and the attacker's specific behavior compose, (2) whether system noise itself is sufficient to make \rscss contention visible, and (3) what impact system noise has on security overall. 

\subsection{Cache-Set Randomization with a Victim Cache}

Another simplification is to omit the complexity of cache skewing and extend cache-set randomization like CEASER~\cite{Qureshi18} with a fully associative \vc. As before, the \vc will decouple evictions being observed from contention in the randomized cache and thus add noise proportional to the \vc size and help reduce re-key intervals. One additional drawback is the smaller number of effective cache sets in this design, which may allow the attacker to exhaustively build eviction sets for all the cache sets and more easily find patterns of interest. More importantly, once the attacker has crafted one eviction set for one set in the cache, the attacker can use these to accurately create contention in the randomized, set-associative cache that will more reliably flush the \vc and help any profiling or attack further on. 

\subsection{Randomized (Skewed) Cache with a Set-Associative Victim Cache}

As \autoref{fig:vcnoise} shows, the noise level introduced by the \vc increases proportional to its size. As this can lead to smaller re-keying rates, a larger \vc is desirable. However, there is a practical size limit for fully associative (victim) caches due to power and area constraints. To yet realize large \vc sizes, a more aggressive design variant is to replace the fully associative \vc with a set-associative \vc that, like the randomized (skewed) main cache, uses a secret mapping to derive the set index. This randomized, set-associative \vc would still be smaller than the randomized main cache and could as well be subject to re-keying. Interestingly, if the main cache uses pure set-randomization like CEASER, a randomized, set-associative \vc adds a second-level mapping similar to cache skewing, but without the necessity to touch the baseline cache's set-associative structure. It seems a relevant question if this two-level design is indeed equivalent to \rscsp or has different security properties and, e.g., allows for new attacks.

\section{Comparison}
\label{sec:comparison}

The objective and design of \cc bears some similarity to Mirage~\cite{saileshwar2021mirage}: both aim to mimic the behavior of a fully-associative cache with random replacement and extend the concept of \rscsp. Moreover, the reinsertion policy of \cc behaves similar to cuckoo relocation as presented for Mirage, but serves a different purpose: \cc uses reinsertion to increase effective associativity, whereas Mirage implements cuckoo relocation to reduce the likelihood of conflicts in the skewed, over-provisioned tag array.

However, there are also some major differences between \cc and Mirage: Mirage is designed to make global replacement decisions as in a fully-associative cache, whereas \cc uses \rscss reinsertion to obtain eviction patterns that look similar to fully associative caches. Moreover, \cc features lower area overheads as it does not require indirection pointers nor over-provisioning of the tags, but may introduce higher cache activity from reinsertion. Both \cc and Mirage show low overheads of $<$1\% and 2\% on spec2017, respectively. Last, \cc introduces the concept of victim caches to \rscsp to hide contention in the cache eviction pattern and stage re-insertions. As highlighted in \autoref{sec:variants}, this victim cache is a versatile tool that thus may also be used to extend Mirage and improve its properties and trade-offs.

\section{Conclusion}
\label{sec:conclusion}

Recent analysis of cache randomization has demonstrated new approaches, e.g., \ppp, to more efficiently learn how addresses map to cache sets and calls for more frequent re-keying to maintain side-channel resilience. However, increasing the re-keying rate comes with higher performance overheads and implementation cost. As a result, this work presented \cc to make cache-set randomization more robust and facilitate reduced re-keying intervals. With the aim to mimic FA caches with random replacement, \cc extends \rscsp with a fully associative victim cache (\vc) and automatic reinsertion. This additional \vc hides contention occurring in the Randomized Skewed Caches (\rscss) by decoupling the \rscss contention from evictions being observed by system users. More importantly, \cc leverages the multiple mappings available in skewed caches to automatically reinsert lines moved to the \vc back into the \rscss in a potentially different division (with a different mapping). This automatic reinsertion mechanism is designed to revert the original \rscss contention and pick an alternative eviction candidate and seeks to obtain eviction patterns that are similar to eviction patterns of fully-associative (FA) caches with random replacement. Thus, \cc can resist fine-grained contention-based attacks and reduce its attack surface to cache utilization channels as in FA caches with rand. replacement. We evaluated the performance of \cc in gem5 showing overheads of $<1\%$ and highlighted the versatility of the \vc in alternative designs to increase side-channel resilience and reduce re-keying rates in randomized caches.

\balance 

\bibliographystyle{IEEEtranSN}
\bibliography{chameleon_cache, spark}

\end{document}